%% file: proceedings.tex
\documentstyle[graphicx,latexsym,multicol]{aipproc}
\include{macro}

\begin{document}

\title{\rxte\ Studies of Cyclotron Lines in Accreting Pulsars}

\author{W.A. Heindl$^*$, W. Coburn$^*$, D.E. Gruber$^*$,
M. Pelling$^*$, R.E. Rothschild$^*$
P. Kretschmar$^{\dagger, \ddagger}$, I. Kreykenbohm$^{\dagger}$, 
J. Wilms$^{\dagger}$,
K. Pottschmidt$^{\dagger}$, and R. Staubert$^{\dagger}$}

\address{$^*$Center for Astrophysics and Space Sciences, University of
California San Diego, La Jolla, CA, 92093, USA\\$^{\dagger}$Institut
f\"{u}r Astronomie und Astrophysik -- Astronomie, Waldh\"{a}user
Str. 64, D-72076 T\"{u}bingen, Germany\\$^{\ddagger}$INTEGRAL Science
Data Centre, Ch. d'Ecogia 16, 1290 Versoix, Switzerland}

\maketitle

%\begin{abstract}
%In this poster, we summarize the \rxte\ measurement of CRSFs in eight
%accreting pulsars.  
%foobar
%\end{abstract}

\section{Introduction}

Cyclotron lines in accreting X-ray pulsar spectra result from the
resonant scattering of X-rays by electrons in Landau orbits on the
intense (\aprx $10^{12}$\,G) magnetic fields near the neutron star
poles.  For this reason they are known as cyclotron resonance
scattering features (CRSFs).  Because Landau transition energies are
proportional to field strength ($\rm E_{cyc}=12$\,keV approximately
corresponds to $B = 10^{12}$\,G), CRSF energies give us our most
direct measures of neutron star magnetic fields.  Other line
properties, such as depths, widths, and presence of multiple
harmonics, are strongly dependent on the details of the geometry and
environment at the base of the accretion column.  CRSFs therefore give
us a sensitive (if difficult to interpret) diagnostic of the accretion
region.

In this paper, we summarize the \rxte\ measurements of CRSFs in 8
accreting pulsars.  The wide bandpass and modest resolution of the
\rxte\ instruments make them ideal for measuring these generally broad
features.  In particular, the high energy response of HEXTE provides
a window for discovery of new lines not detectable with
proportional counters such as \ginga\ or the PCA alone.

Some highlights of this work are: 
\begin{list}{$\bullet$}%
{\setlength{\topsep}{0.in}%
 \setlength{\parsep}{0.in}%
 \setlength{\partopsep}{0.in}%
 \setlength{\itemsep}{0.5ex}}%
	\item{New lines in the well known
	pulsars Cen~X-3 and 4U~1626-67 \cite{hei99b,San98,Orl98}. }
	\item{The discovery of more than two cyclotron line harmonics in
	4U~0115+63 \cite{Hei99}}.
	\item{The \rxte\ picture of the correlation between cyclotron
	line energy and width.}
\end{list}

\section{Observations and Analysis}

The results discussed here are based on observations made with the
Proportional Counter Array (PCA) \cite{jah96} and High Energy X-ray
Timing Experiment (\hxt) \cite{rot98} on board \rxte.  The PCA is a
set of 5 Xenon proportional counters, while \hxt\ consists of two
arrays of 4 NaI(Tl)/CsI(Na) phoswich scintillation counters. The PCA
and \hxt\ share a 1\degree\ field of view.
\begin{figure}
\centerline{\includegraphics[height=1.8in,bb= 130 95 517 503]{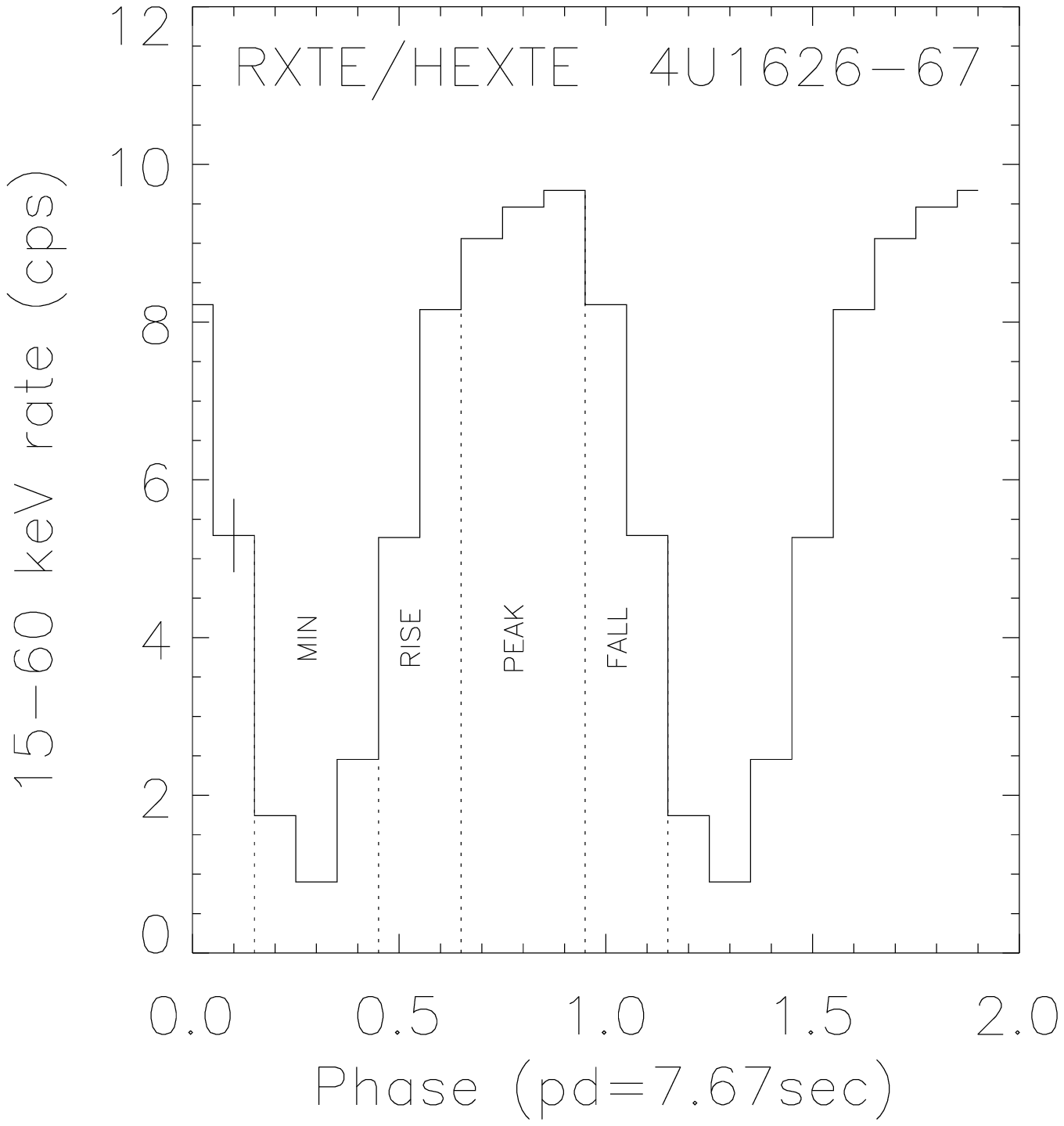}
\hspace{0.05in}\includegraphics[height=1.8in]{prs_forMunich98.ps}
\hspace{0.05in}\includegraphics[height=1.8in,bb=103 104 543 518]{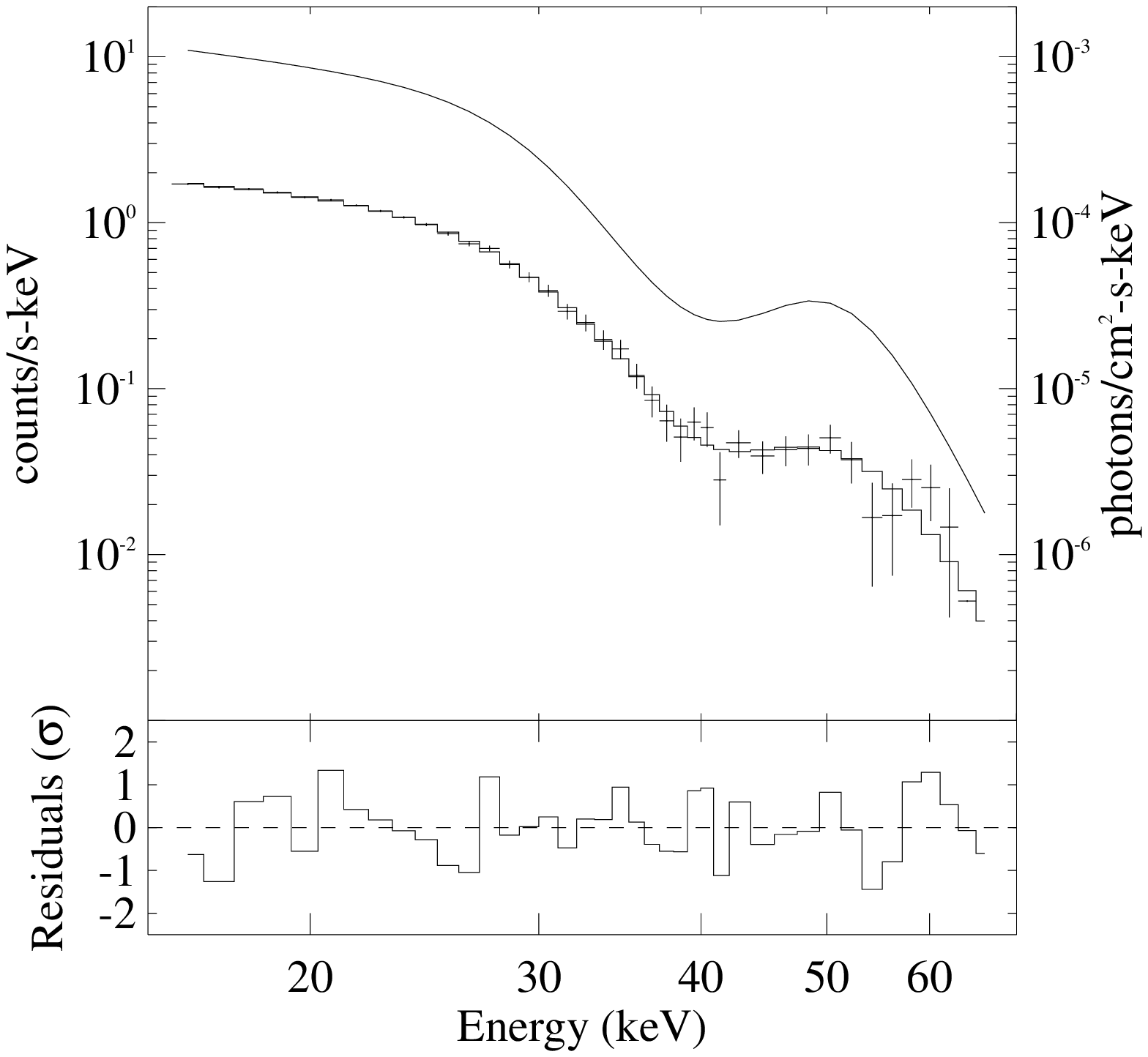}}
\vspace{0.075in}
\centerline{\includegraphics[height=1.8in,bb=104 84 523 512]{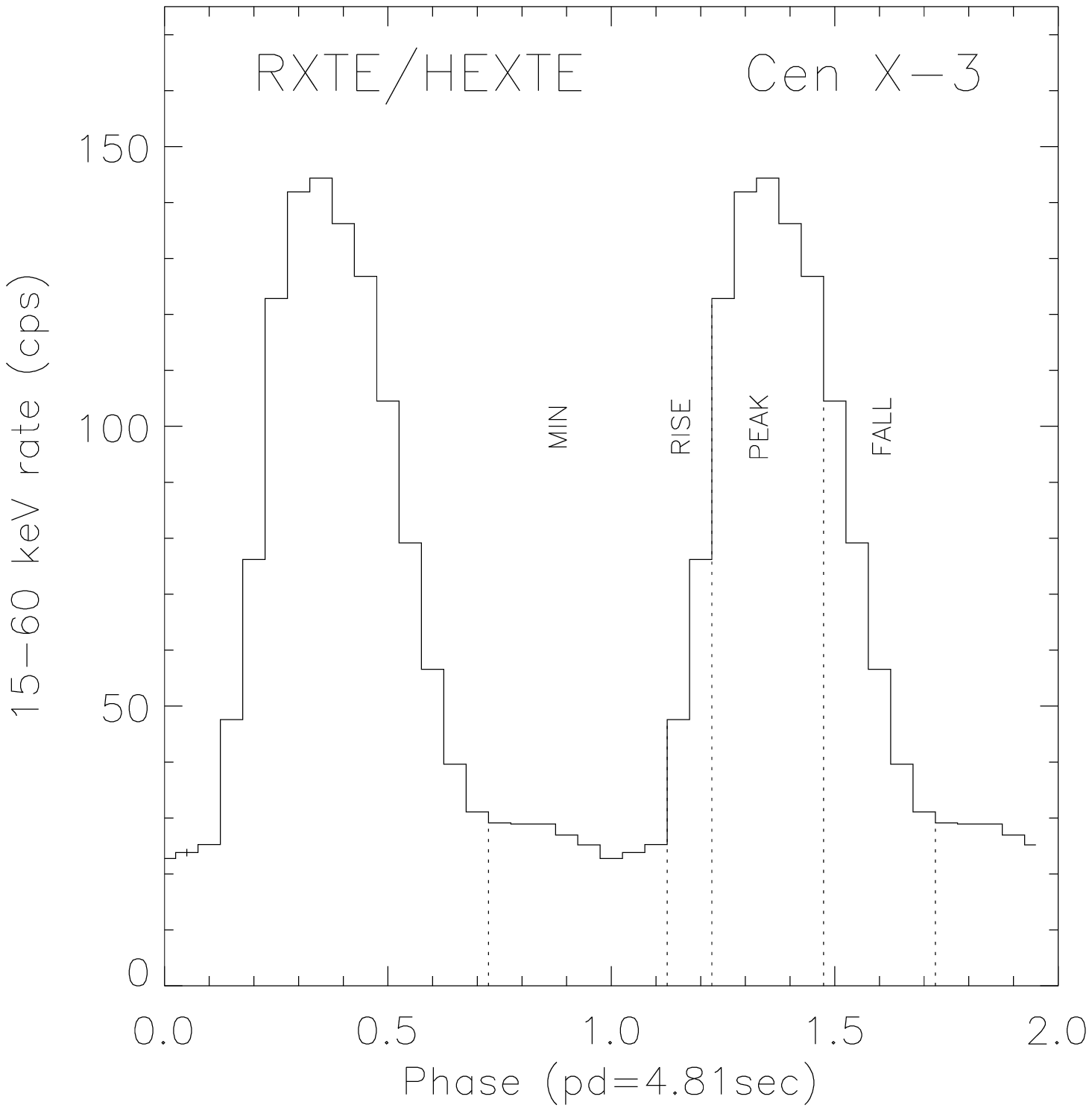}
\hspace{0.05in}\includegraphics[height=1.8in]{prs_cenx3_munich98.ps}
\hspace{0.05in}\includegraphics[height=1.8in,bb=128 132 529 518]{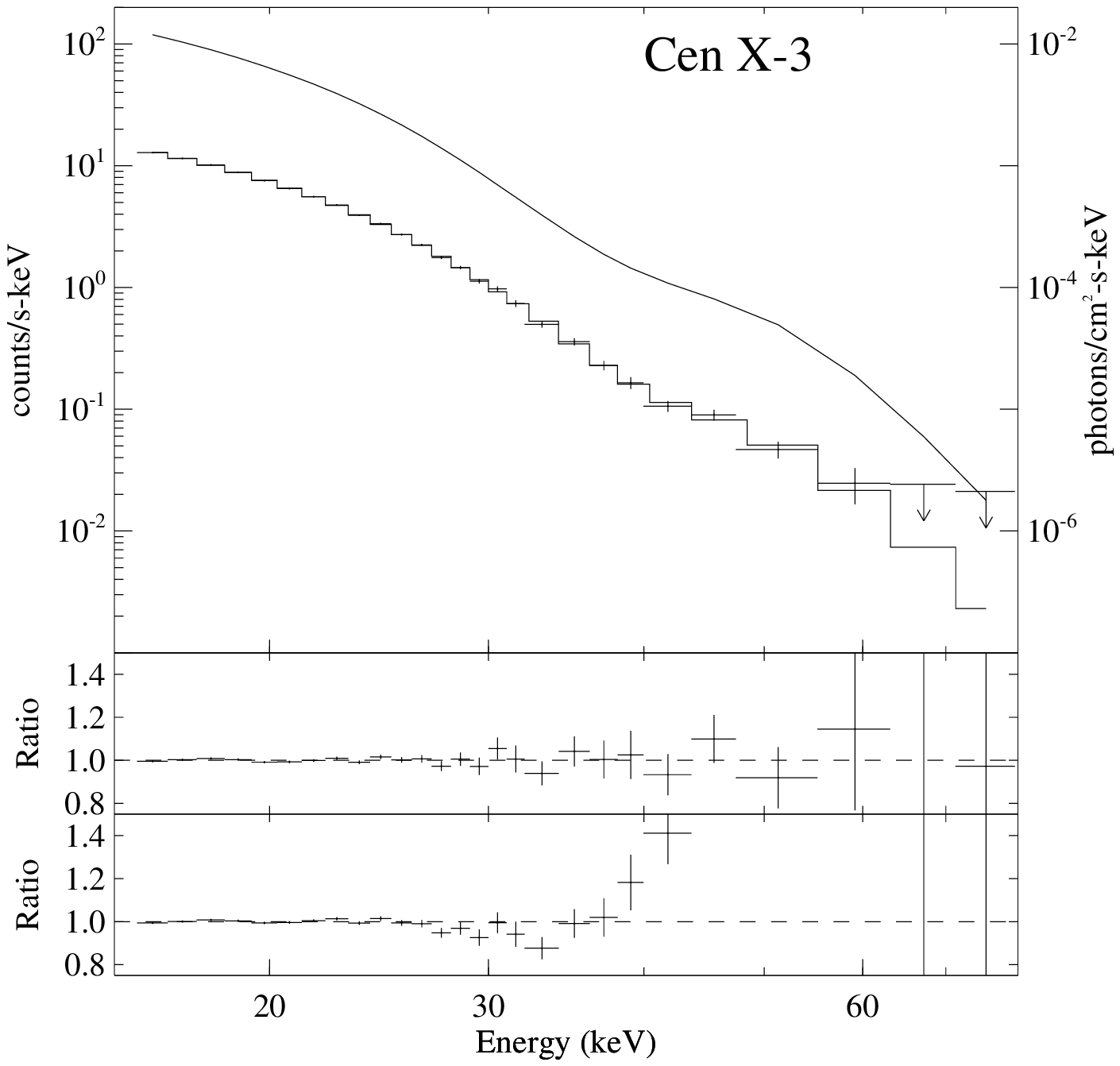}}
\vspace{0.075in}
\caption{\label{fig:cen}\rxte\ observations of (Top) 4U~1626-67 and
(Bottom) Cen X-3 .  Left: HEXTE light curves folded on the pulse
period. %Four phase bands used in spectral analyses are indicated.
Middle: The counts spectra from the four phase intervals.  The CRSFs
are apparent as inflections in the spectra near 30--40~keV.  Right:
the inferred incident spectrum from the peak phase, HEXTE count
spectrum with model, and ratios to models with (and for Cen X-3
without) a cyclotron line.}
\end{figure}

Figures~\ref{fig:cen} and \ref{fig:allsix} show the measured count
spectra together with model fits and inferred incident spectra for
eight accreting pulsars.  To emphasize the presence of cyclotron
lines, residuals to fits made without lines are also shown.  In
general, these residuals show a dip-like structure at the line energy
and then a gross underprediction of the continuum above the line. This
is caused by the better statistics on the low side of the feature
forcing the model to fit the falling edge of the line.  Because no
adequate prediction of accreting pulsar continua exists, we employ
empirical continuum models. These models (high energy cut-off power
law (HECUT), Fermi-Dirac cut-off times a power law (FDCO), and
Negative and Positive power law Exponential (NPEX); see \cite{kre98})
have been successful in fitting pulsars with no cyclotron lines. Only
when none of these continuum models provided an acceptable fit, did we
allow cyclotron line(s).  We modeled the cyclotron lines with a simple
Gaussian optical depth profile.

\section{Results and Discussion}

Table~\ref{tab:fits} summarizes the cyclotron line parameters in eight
accreting pulsars.  In general, pulsar spectra (including line parameters) vary
with pulse phase.  For this reason, cyclotron lines are often best
measured in spectra from limited pulse phase ranges. This is called
``pulse phase spectroscopy''.  Table~\ref{tab:fits} indicates the
phase relative to the pulse profile for which the given
parameters apply.
\begin{table}
\caption{\label{tab:fits}Summary of \rxte\ Cyclotron Line
Measurements. 4U~0115+63 and Vela X-1 require multiple CRSFs in their
spectra.  The surface B-fields assume a gravitational redshift of 0.3
for the emitting region.}
\label{t_crsf}
\begin{tabular}{lccccc}%
%\hline 
Source  	&    Energy   &    Sigma   &   Optical   & Pulse Phase  & Surface B Field\\
		&  (keV)      &   (keV)    &   Depth     &
		& ($ 10^{12}$\,G)  \\\hline
4U 0115+63	& $\rm 12.40^{+0.65}_{-0.35}$	&$\rm 3.3^{+0.1.9}_{-0.4}$&$\rm 0.72^{+0.10}_{-0.17} $& $\rm2^{nd}$ Fall & 1.4\\
		& $\rm 21.45^{+0.25}_{-0.38}$	&$\rm 4.5^{+0.7}_{-0.9} $ &$\rm 1.24^{+0.04}_{-0.06} $& $\rm2^{nd}$ Fall & \\
		& $\rm 33.56^{+0.70}_{-0.90}$	&$\rm 3.8^{+1.5}_{-0.9}$& $\rm 1.01^{+0.13}_{-0.14} $& $\rm2^{nd}$ Fall & \\
4U 1907+09	& $\rm 19.7 \pm 0.1$		& $\rm2.6 \pm 0.1$	& $\rm0.87 \pm 0.05$	& Average  & 2.2\\ 
4U 1538-52	& $\rm 21.3\pm0.3 $		& $\rm3.05 \pm0.20 $	& $\rm0.86 \pm 0.07$	& Average  & 2.3\\
Vela X-1	& $\rm 23.7^{+0.4}_{-0.3}$	& 5 (fixed)		& $0.29^{+0.03}_{-0.04}$& Main Pulse & 2.3\\
		& $\rm 59.7\pm3.7$		& $12.6\pm0.8$		& $1.41^{+0.67}_{-0.61}$& Main Pulse & \\
Cen X-3		& $\rm 31.8\pm 0.3$		& $7.5\pm 0.9$		& $0.77^{+0.16}_{-0.11}$ & Peak & 3.5\\
GX 301-2	& $\rm 42.9^{+0.9}_{-2.6}$	& $10.0^{+1.9}_{-2.3}$  & $0.8^{+0.7}_{-0.3}$	&  Average & 4.8 \\
4U 1626-67	& $\rm 39.3^{+1.9}_{-1.1}$	& $6.4\pm 0.8$ & $2.3^{+0.6}_{-0.4}$ & Peak & 4.4\\
Her X-1		& $\rm 41.0\pm1.0$		& $9.8\pm0.5$ & $1.84\pm0.05$ & Average& 4.6 \\ \hline
\end{tabular}
\end{table}

\noindent{\bf New Lines in Cen~X-3 and 4U~1626-67\hspace{1.5em}}
Cen~X-3 and 4U~1626-67 are two of the earliest known accreting
pulsars; however, it is only recently that CRSFs were discovered in
their spectra \cite{hei99b,Ddf99}.  Figure~\ref{fig:cen} shows folded
light curves, phase resolved count spectra, and fits to the peak phase
spectra. In Cen X-3, the CRSF energy moves by about 20\% with phase.
This variation has been modeled as resulting from an offset of the
magnetic dipole moment from the center of the neutron star
\cite{Bur99}.  In both sources, the CRSF appears strongest in the
``peak'' and ``fall'' spectra. This may also be related to an offset
dipole.

\begin{figure}
\centerline{\includegraphics[width=2.5in,bb=121 133 529 517]{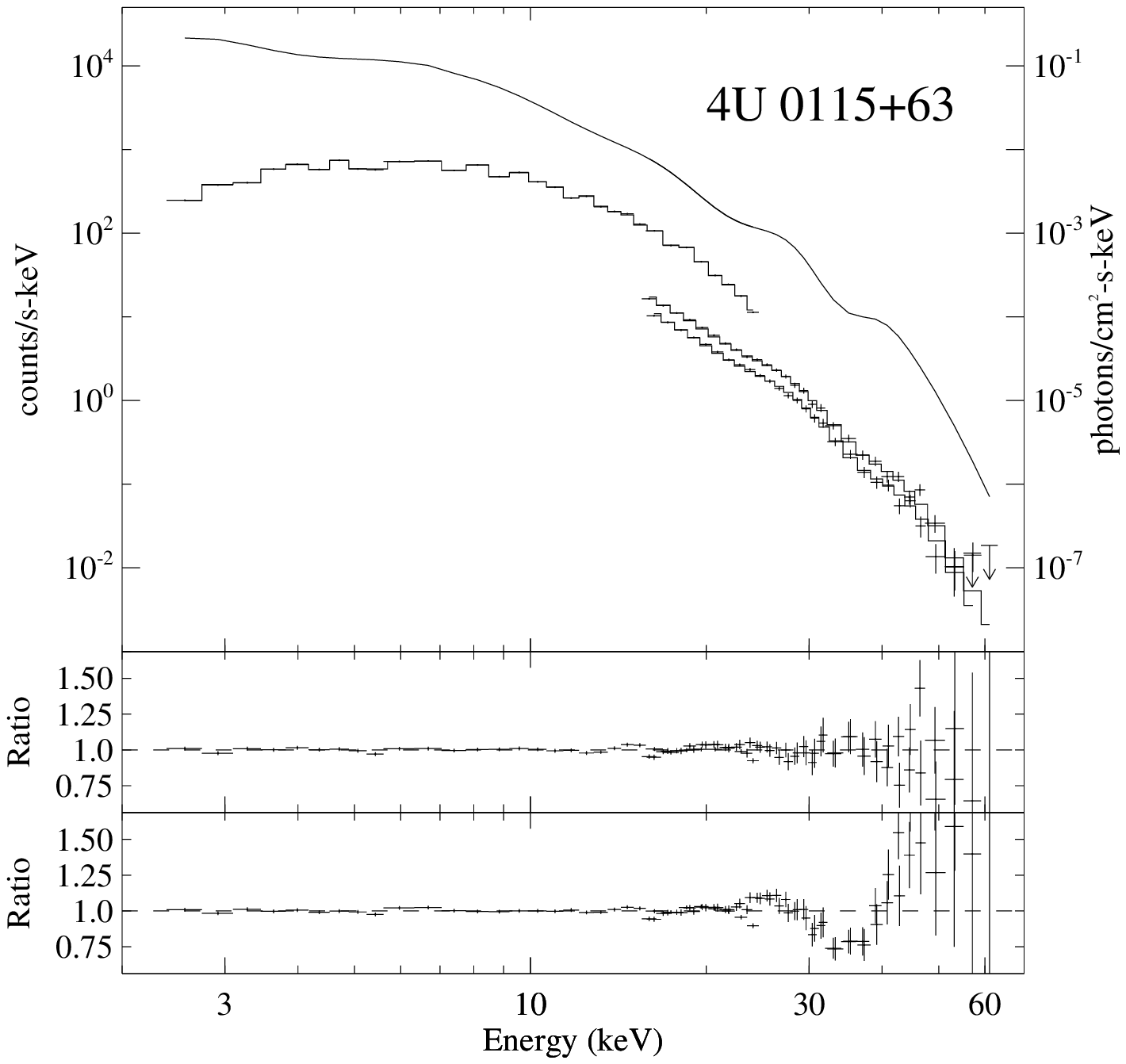}
\hspace{0.25in}\includegraphics[width=2.5in,bb=128 133 529 517]{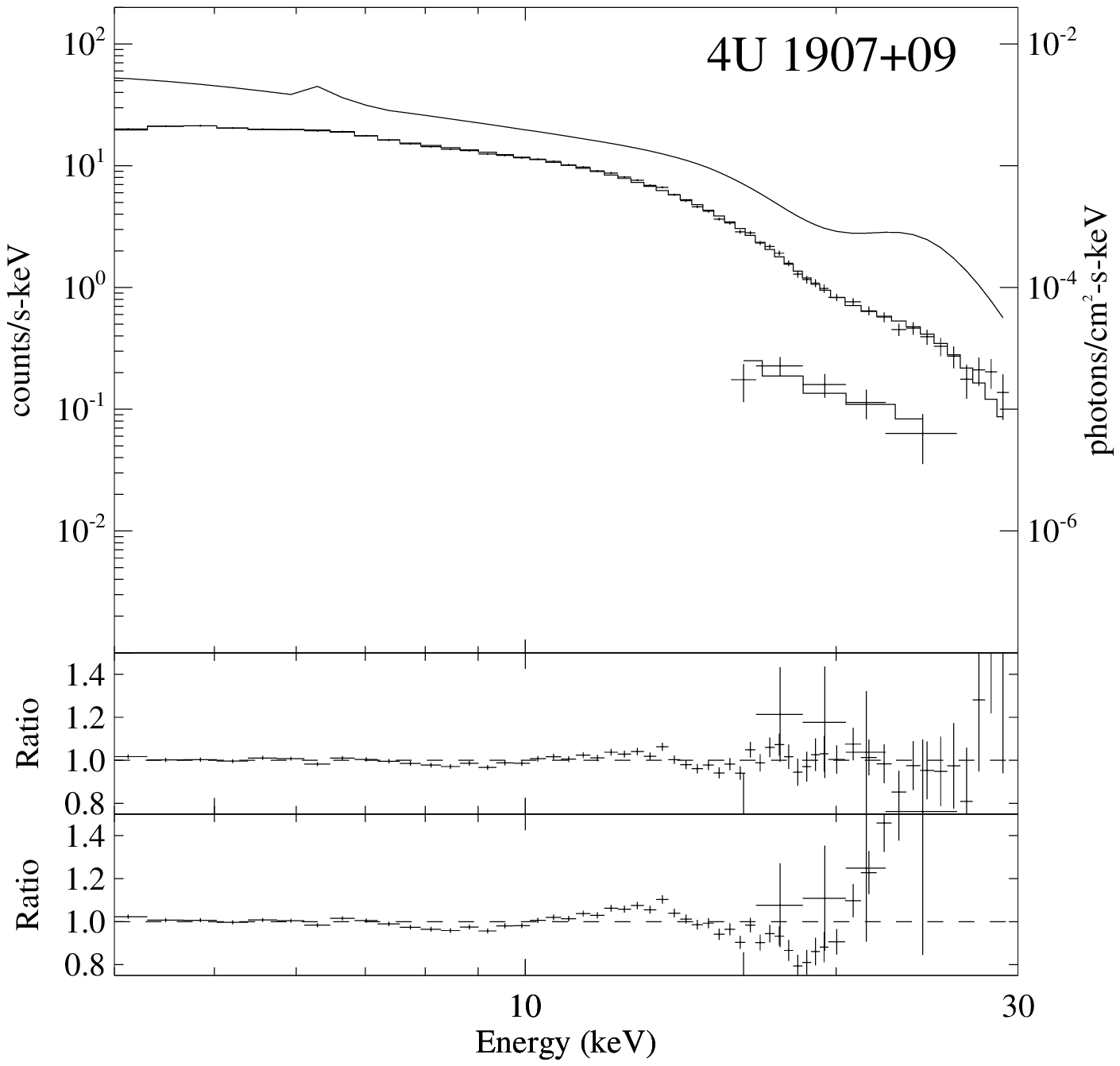}}
\vspace{0.2in}
\centerline{\includegraphics[width=2.5in,bb=128 134 529 517]{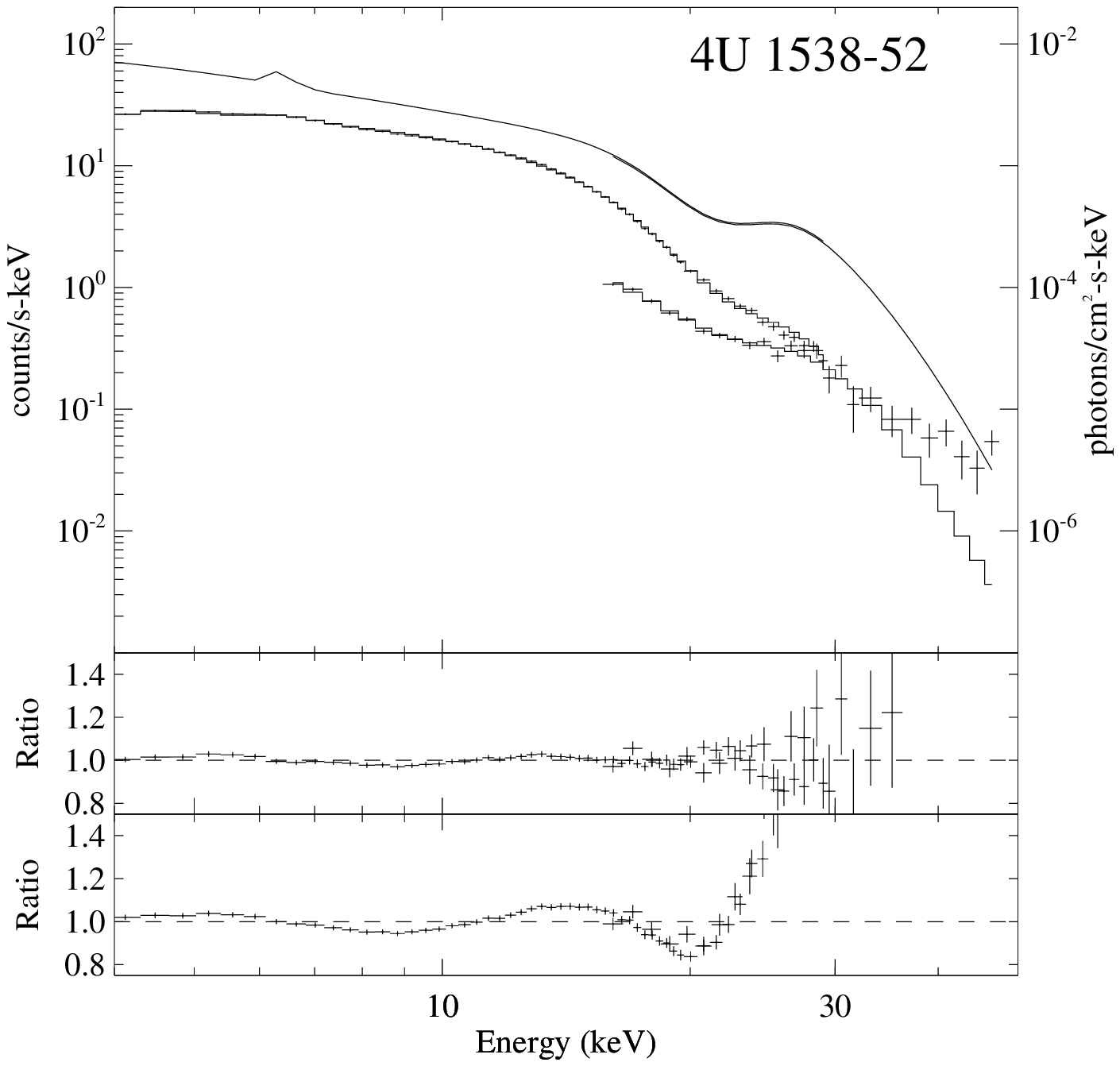}
\hspace{0.25in}\includegraphics[width=2.5in,bb=128 134 529 517]{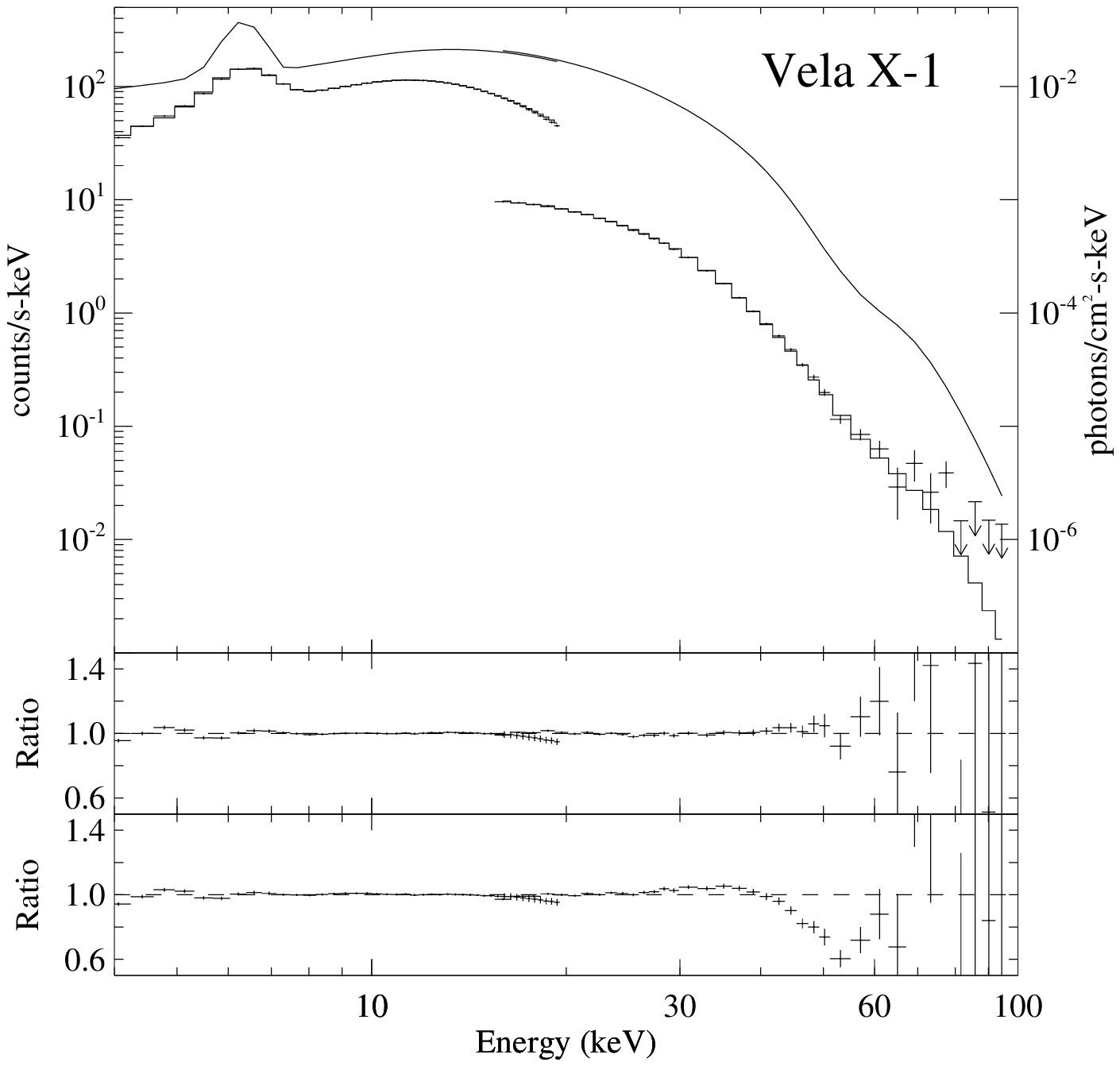}}
\vspace{0.2in}
\centerline{\includegraphics[width=2.5in,bb=128 134 529 517]{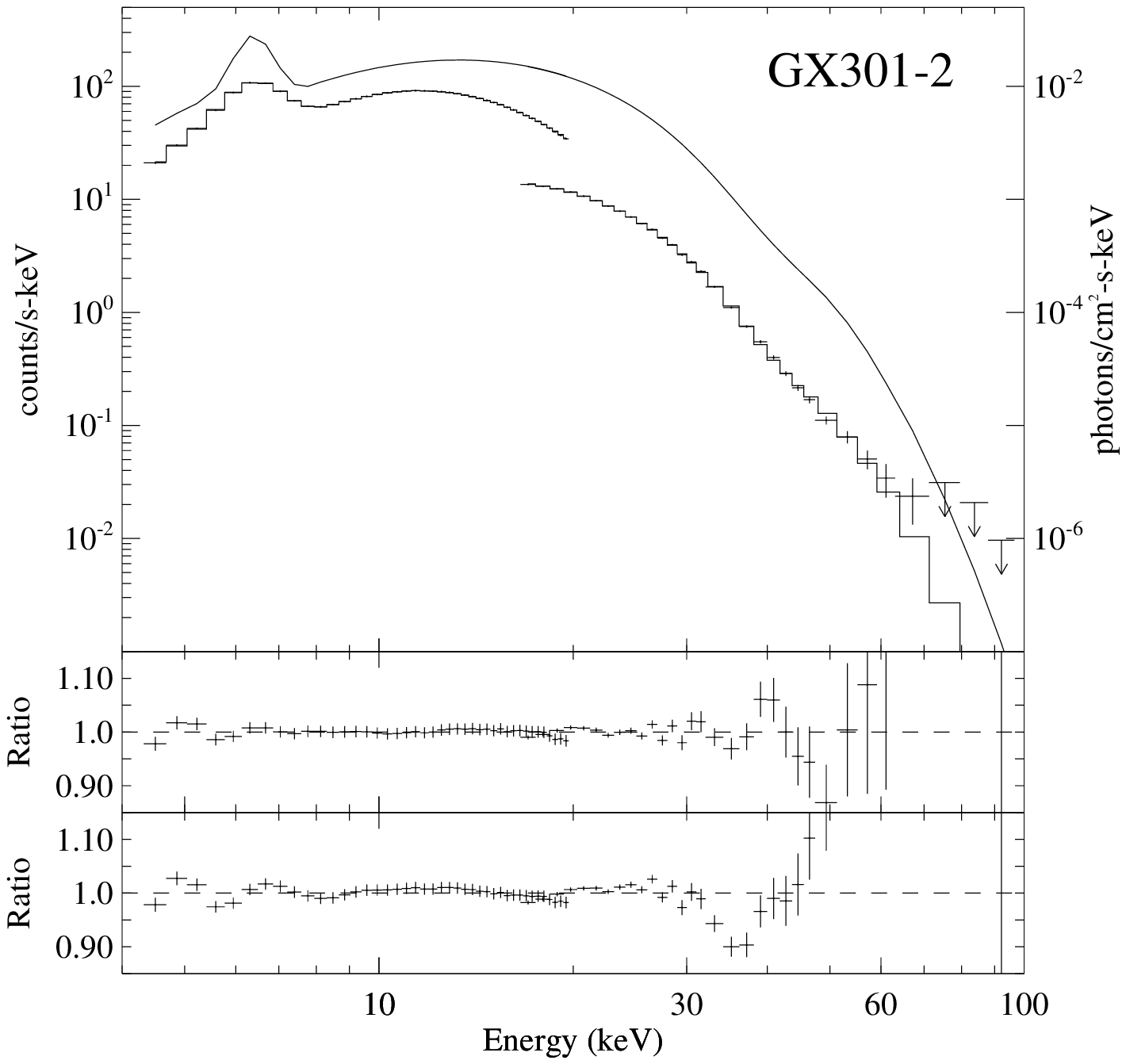}
\hspace{0.25in}\includegraphics[width=2.5in,bb=128 134 529 517]{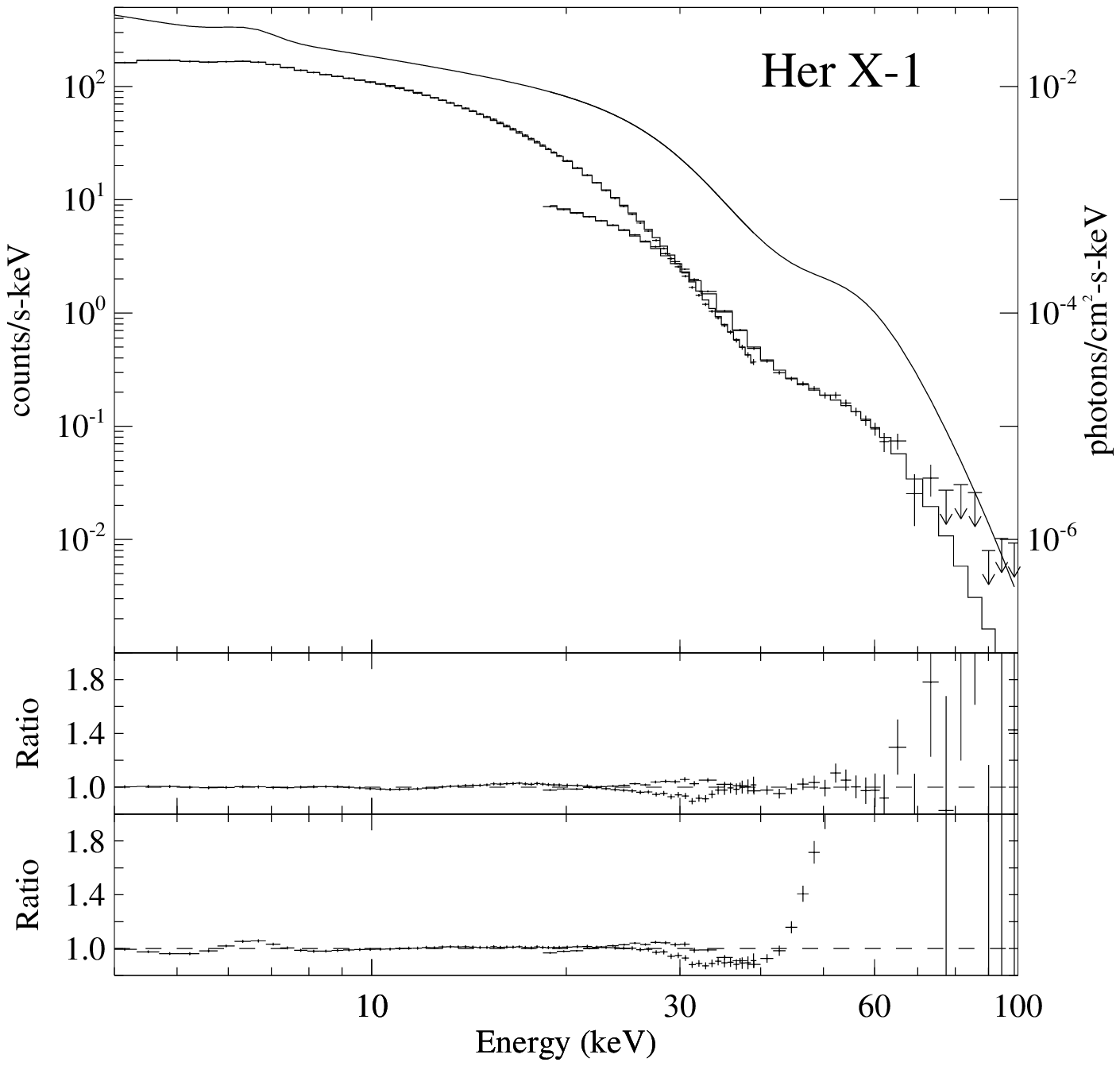}}
\caption{\label{fig:allsix} \rxte\ spectra of six accreting
pulsars. Top panels show inferred incident spectra (solid lines) and
PCA and HEXTE counts spectra. Middle panels show the ratio of the best
fit model, which includes a CRSF, to the data.  Bottom panels show
this ratio for a model which has no CRSF.  In the case of 4U~0115+63,
which has three lines in this spectrum, we show residuals to models
with three and two lines included. For line parameters, see
Table~\ref{tab:fits}.  }
\end{figure}

\noindent{\bf 4U~0115+63: Multiple Cyclotron Line
Harmonics\hspace{1.5em}} 4U~0115+63 is a transient accreting pulsar in
an eccentric orbit with a massive star, a so-called ``Be X-ray
binary''.  It was also the first pulsar to show two cyclotron line
harmonics \cite{wsh83}. 4U~0115+63 underwent a two month long
outburst in 1999 February--April.  Observations made with \rxte\ and
\sax\ reveal for the first time more than two harmonics in a single
pulsar \cite{hei99c,san99a} (see Fig.~\ref{fig:allsix}).  

\noindent{\bf Correlation between Line Energy and Width\hspace{1.5em}}
The width of a CRSF depends on the temperature of the emitting region
(kT), the line energy ($\omega_B$) and the viewing angle ($\theta$)
with respect to the magnetic field as: $\Delta \omega \propto \omega_B
(kT/mc^2)^{1/2} cos \theta$ \protect\cite{Mes85}.  This predicts that
a correlation between line energy and width should exist, provided
that viewing angles and plasma temperatures do not vary too greatly
from source to source. These \rxte\ results, as well as measurements
by \sax \protect\cite{Ddf99}, show a strong correlation, which
supports this picture.
\begin{figure}
\centerline{\includegraphics[angle=90,width=4in]{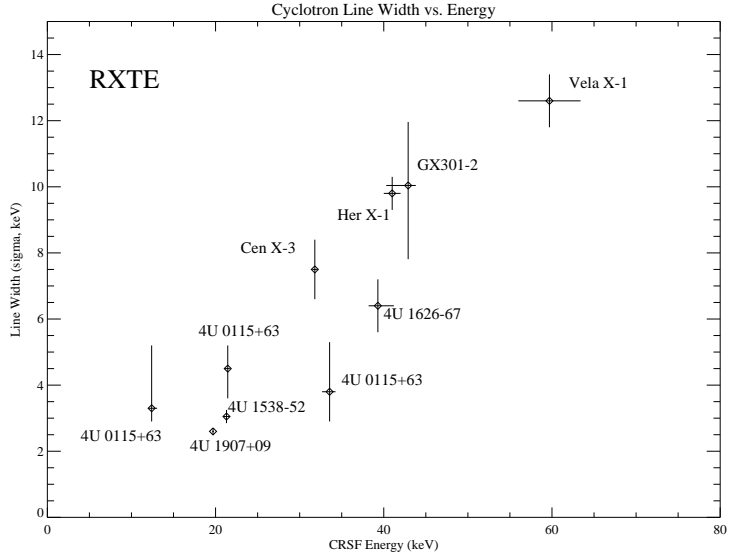}}
\caption{ CRSF width versus line energy for all the eight pulsars
discussed here. A positive correlation between energy and with is
apparent.  Note that for the three cyclotron lines in 4U~0115+63, the
correlation does not appear to apply. }
\end{figure}

%\small
%\begin{multicols}{2}
{\bibliographystyle{amsplain}
\bibliography{proceedings}}
%\end{multicols}
%\begin{thebibliography}{}
%\thebibliography
%\end{thebibliography}

\end{document}

%% file: macro.tex
\newcommand{\ltsim}{\lower.5ex\hbox{$\; \buildrel < \over \sim \;$}}
\newcommand{\gtsim}{\lower.5ex\hbox{$\; \buildrel > \over \sim \;$}}

\newcommand{\aprx}	{\mbox{$\sim$}}
\newdimen\digitwidth
{\catcode`?=\active
        \gdef\setmathspace   {
            \setbox0=\hbox{\rm0}
            \digitwidth=\wd0
            \catcode`?=\active
            \def?{\kern\digitwidth}
        }
}

\def\ginga{{\it Ginga}}

\def\sl2{{Spacelab~2}}

\def\heao3{{\it HEAO 3}}

\newcommand{\rxte}{{\em RXTE}}
\newcommand{\sax}{{{\em Beppo}SAX}}

\newcommand{\hxt}{HEXTE}

\def\a0{\mbox{A~0535+26}}
\def\grs1915{\mbox{GRS~1915+105}}
\def\GX354{\mbox{GX354+0}}

\def\j1744{\mbox{GRO~J1744-28}}

\def\degree{\hbox{$^\circ$}}